# Inconsistency in Fermi's probability of the quantum states

Mofreh R. Zaghloul

Department of Physics, College of Sciences, United Arab Emirates University,
Al-Ain, 17551, UAE. E-mail: m.zaghloul@uaeu.ac.ae

## Abstract

We point out an important hidden inconsistency in Fermi's probability of the quantum states that engendered inconsistent/inaccurate equations-of-state extensively used in the literature to model nonideal plasma systems. The importance of this amendment goes beyond rectifying our comprehension and foundation of an important physical problem to influencing contemporary research results.
.

## Introduction

The paradox of the divergence of atomic internal partition function (IPF) for a hypothetical isolated atom in an infinite space, due to the existence of an infinite number of bound states below the ionization limit, is well recognized and well documented in the literature (see for example Refs [Fermi 1924, Olsen 1961, Strickle 1966, Zel'dovich 1966, Blinder 1995]). This paradox is to a certain extent deceiving, since in reality an isolated atom in an infinite space would never exist. Rather, in reality, atoms are always contained in a gas of finite density. For systems of finite densities, there have been numerous attempts to consider the effect of the environment on the truncation of the infinite sum of quantum states and the establishment of a finite internal partition function.

In a historic work by Fermi (see Ref. [Fermi 1924]), the equilibrium among different quantum states of the same atom, whose energies are $w_1$, $w_2$, ...,$w_r$ was considered assuming *equal statistical weights* of the quantum states and the probability that an atom may be in the $i^{th}$ state was, therefore, written as

$$p_i = C e^{-\frac{w_i}{KT}}, \tag{1}$$

where $T$ is the absolute temperature, $K$ is the Boltzmann constant and $C$ is a constant determined by the normalization condition $\sum_i p_i = 1$.

In his work, Fermi showed that this does not hold true by considering the case of a hydrogen atom, for which the energy $w_i$ of the $i^{th}$ state is given by $w_i = -\frac{R}{i^2}$ where $R$ is the Rydberg energy=13.6 eV. For such a case the probability of the $i^{th}$ state, according to Eq. (1), would be

$$p_i = C e^{\frac{R}{i^2 KT}}, \tag{2}$$

Accordingly, the constant $C$ is given by

$$C = \frac{1}{\sum_i e^{\frac{R}{i^2 KT}}}. \tag{3}$$

As the sum in the denominator of $C$ in Eq. (3) is divergent, one would obtain the wrong answer for $C$, namely $C = 0$.

In an attempt to circumvent this problem Fermi studied the thermodynamic equilibrium between quantum states taking the finite volume of atoms into account. Considering a mixture of $n_1$ atoms in the first quantum state, $n_2$ atoms in the second, …, $n_r$ atoms in the $r^{th}$ quantum state and assigning $v_1, v_2, …, v_r$ to the respective volumes of $r$-quantum states of the atom under consideration he started from the Helmholtz free energy, $F$, given formally by

$$F = U - TS \tag{4}$$

where $U$, $S$, and $T$ are, respectively, the total energy, the entropy and the absolute temperature of the mixture under consideration. In Eq. (4), $U$ is the sum of the kinetic and internal energies of the atoms. Thus, $U$ was expressed as

$$U = \frac{3}{2} n K T + \sum n_i w_i \tag{5}$$

where

$$n = n_1 + n_2 + ... + n_r \tag{6}$$

is the total number of atoms under consideration.

To calculate the entropy, Fermi used the van der Waals equation of state of the mixture without the term $A/v^2$, namely $p(V-b) = nKT$, where $V$ is the volume of the mixture and the excluded volume $b$ is given by

$$b = \frac{1}{2n} \sum_{ik} n_i \, n_k \, \left( \sqrt[3]{v_i} + \sqrt[3]{v_k} \right)^3 \tag{7}$$

As a result, the entropy of the mixture was expressed as

$$S = \frac{3}{2} n K \int \frac{dT}{T} + n K \int \frac{dV}{V-b} + const$$
$$= \frac{3}{2} n K \, \ln T \, + n K \, \ln(V-b) - K \sum n_i \ln n_i \tag{8}$$

The Boltzmann constant $K$ in the far most term in the right hand side of Eq. (8) was missed in Fermi's original article and has been inserted herein for dimensional correctness of the equation.

The free energy of the mixture can, therefore, be written as

$$F = \frac{3}{2} n K T + \sum_i n_i w_i - n K T \left\{ \frac{3}{2} \ln T + \ln V + \ln(1 - \frac{b}{V}) - \sum_i \frac{n_i}{n} \ln n_i \right\} \tag{9}$$

Assuming that $b$ is very small in comparison to $V$, one can write

$$F \cong \frac{3}{2} n K T + \sum_i n_i w_i - n K T \left\{ \frac{3}{2} \ln T + \ln V - \frac{b}{V} - \sum_i \frac{n_i}{n} \ln n_i \right\} \tag{9'}$$

Equation (9) or (9') as it stands above was introduced to represent the free energy of an ideal gas, in addition to a separable configurational component, $F_{conf}$, given by $F_{conf} = -n K T \ln(1 - \frac{b}{V}) \approx n K T \frac{b}{V}$. Equation (9) or (9') can, therefore, be formally written as

$$F = \frac{3}{2} n K T + \sum_i n_i w_i - n K T \left\{ \frac{3}{2} \ln T + \ln V - \sum_i \frac{n_i}{n} \ln n_i \right\} + F_{conf} \tag{10}$$

At equilibrium, the free energy is minimized and the following minimization condition has to be satisfied,

$$\frac{\partial F}{\partial n_1} = \frac{\partial F}{\partial n_2} = ... = \frac{\partial F}{\partial n_r} = const. \tag{11}$$

From this condition Fermi derived for the occupation numbers $n_i$ the following expression

$$n_i = C\, e^{-\frac{w_i}{KT}}\, e^{\frac{\partial n\ln(1-b/V)}{\partial n_i}} \approx C\, e^{-\frac{w_i}{KT}}\, e^{-\frac{1}{V}\frac{\partial nb}{\partial n_i}} = C\, e^{-\frac{w_i}{KT}}\, e^{-\sum_k \frac{n_k}{V}\left(\sqrt[3]{v_i}+\sqrt[3]{v_k}\right)^3} \tag{12}$$

For a general form of the configurational component of the free energy, the above expression can be generalized to

$$n_i = C\, e^{-\frac{w_i}{KT}}\, e^{-\frac{1}{KT}\frac{\partial F_{conf}}{\partial n_i}} \tag{13}$$

Simultaneous satisfaction of Eq. (6) and the normalization condition, $\sum_i p_i = 1$, requires that $C = n/Q_{int}$ where $Q_{int}$ is the sum over all states (internal partition function) given by

$$Q_{\text{int}} = \sum_i e^{-\frac{w_i}{KT}}\, e^{-\frac{1}{KT}\frac{\partial F_{conf}}{\partial n_i}} \tag{14}$$

Comparing Eq. (12) or Eq. (13) with Eq. (1) one sees that the difference resides in the factor $e^{-\sum_k \frac{n_k}{V}\left(\sqrt[3]{v_i}+\sqrt[3]{v_k}\right)^3}$ or the factor $e^{-\frac{1}{KT}\frac{\partial F_{conf}}{\partial n_i}}$ for a general form of the configurational component.

Considering the "low excitation approximation" and assuming that, for a hydrogen atom, the volume can be expressed in the form of a sphere of radius equal to the semi-major axis of a Keppler ellipsoid, the volume, $v_i$ is given by $Ai^6$ where $A$ is a constant given approximately by $A=5\times10^{-25}$, the above factor can be reduced to $e^{-Ai^6\frac{n_i}{V}}$, which has been interpreted as a "probability a priori". It was concluded, therefore, that Eq. (2) should be replaced by

$$p_i = C\, e^{\frac{R}{i^2 KT} - Ai^6\frac{n_i}{V}} \tag{15}$$

where $C$ is given by satisfying the constraints $\sum_i p_i = 1$ and Eq. (6) resulting in

$$C = \frac{n}{Q_{\text{int}}} = \frac{n}{\sum_i e^{\frac{R}{i^2 KT} - Ai^6\frac{n_i}{V}}} \tag{16}$$

The sum in the denominator of Eq. (16) is now convergent and the problem of the divergent sum over all quantum states was thought to be solved.

This approach to circumvent the paradox of divergent partition function was adopted latter by Hummer & Mihalas [Hummer 1988] who reintroduced this approach in the form of the occupation probability formalism used in the calculation of the equation of state for stellar envelopes in the Opacity Project (OP). The occupation probability formalism has become most popular, for quenching the divergence of the atomic internal partition functions, in astrophysics after the series of papers by Hummer, Mihalas, Däppen, Nayafonov, and others (see Refs. [Hummer 1988, Mihalas 1988, Nayfonov 1999, Däppen 2000, Potekhin 1996]). However, we find this approach as well as all of its engendered clones to be inconsistent or inaccurate as we explain below.

## The Inconsistency

There are many ways to show the statistical-mechanical-inconsistency and/or the inaccuracy in Fermi's solution of the paradox of the divergence of the atomic internal partition function. Herein, we present a few of these reasons and proofs that are sufficient to explain the inconsistency and/or inaccuracy in Fermi's probability of quantum states.

First of all, the inconsistency in Fermi's treatment is axiomatically recognizable from the assumption of separability of the free energy components (uncoupling of various types of energies) as it appears in Eq. (10). Simply, the separation of the configurational component of the free energy implies that it has no influence on the internal free energy component and, therefore, the expectation that including a separable configurational component could lead to a truncation of the internal partition function is conceptually incorrect because they are independent from each other by assumption,

Secondly, in the determination of the distribution of atoms among various quantum states as presented by Fermi it was implicitly assumed in the derivation of the entropy expression of the gas (Eq. (8)) that the specific heat at constant volume is temperature-independent which is in contradiction with the inclusion of the excitation states and excitation energies in the expressions used for the entropy and energy of the system. In addition to this, one has to remember that the formal thermodynamic (i.e. macroscopic) entropy of a gas following van der Waals equation of state without the term $A/v^2$, namely $p(V-b) = nKT$ is given by

$$S(T,V,n) = n \int \frac{c_v(T')\,dT'}{T'} + nK \int \frac{dV}{V-b} + C'$$

$$= n \int \frac{c_v(T')\,dT'}{T'} + nK \ln(V-b) + C' \qquad (17)$$

$$\approx n \int \frac{c_v(T')\,dT'}{T'} + nK \ln(V) - nK \frac{b}{V} + C'$$

where $c_v$ is the heat capacity at constant volume per molecule and $C'$ is a constant of integration (essentially the entropy at some reference state). The constant of integration $C'$ is independent of $T$ and $V$ but may depend on $n$. Reverting to statistical mechanical results, Fermi used for the constant $C'$ the expression $-K \sum n_i \ln n_i$. An additional term $Kn\left(\ln\left(2\pi mK/h^2\right)^{3/2} + 5/2\right)$, where $m$ is the molecular mass and $h$ is Planck's constant, which was missing in Fermi's expression for the entropy is included in Hummer's and Mihalas treatment [Hummer 1988]. Nevertheless, it has to be clear that reverting to statistical mechanics results to express the constant $C'$, as indicated above, would essentially require that the derived occupations or probabilities of the quantum states must comply with and must satisfy the fundamental statistical thermodynamic relations. In particular, for these probabilities to be statistical mechanically consistent they must satisfy the fundamental relation

$$F = -KT \ln Z_{tot} \tag{18}$$

where $Z_{tot}$ is the total partition function of the mixture. It is basic that the complement of the separability of the free energy components (uncoupling of various types of energies) as adopted in Fermi's formulation (see Eqs (9,10), for example) is the factorizability of the bulk-state partition function and hence the ability to express it in terms of the partition functions of individual molecules, that is

$$F = F_{tran} + F_{int} + F_{conf} = -KT \ln(Z_{trans} \, Z_{int} \, Z_{conf})$$

$$= -KT \ln\left( \frac{(Q_{trans} \, Q_{int} \, Q_{conf})^N}{N!} \right) \equiv -NK \, T \ln\left( \frac{Q_{trans} \, Q_{int} \, Q_{conf}}{N} + 1 \right) \tag{19}$$

where $Z_{trans}$, $Z_{int}$, and $Z_{conf}$ are the translational, internal and configurational factors of the bulk-state partition function while $Q_{trans}$, $Q_{int}$, and $Q_{conf}$ are the corresponding individual molecular factors or partition functions. In deriving the final result in Eq. (19), Stirling's approximation has been used.

It is very simple and straightforward to show that using Fermi's probabilities of the quantum states (Eqs (12 or 13) into his free energy function (Eqs (9 or 10)) can not lead to the fundamental statistical thermodynamic relations (Eqs (19 or 18)) except for forms of $F_{conf}$ that are linearly dependent on the occupations or probabilities of individual quantum states. However, for such a case the factor $e^{\frac{1}{KT} \frac{\partial F_{conf}}{\partial n_i}}$ in Eqs (13) and (14) will be independent of the individual quantum states and, therefore, can be factored out from the sum in Eq. (14) with no guarantee of the truncation of the internal partition function in such a case and the problem of the divergence of the internal partition persists.

Thirdly, it is very simple to show the inconsistency in Fermi's treatment by considering his expression for the entropy (Eq. (8)) which embodies the excluded volume or the configurational component in it. Recalling that the entropy is an extensive thermodynamic property, the entropy expression given by Eq. (8) necessitates that $b$ be independent of the individual excited states. However, the expression given and used for $b$ in Eq. (7) does not satisfy this condition and Fermi's entropy expression violates a basic thermodynamic requirement. On the other hand if one chooses for $b$ an expression that satisfies this requirement, the factor $e^{\frac{1}{V} \frac{\partial nb}{\partial n_i}}$ in Eq. (12) or $e^{\frac{1}{KT} \frac{\partial F_{conf}}{\partial n_i}}$ in Eq. (13) will be independent of the individual quantum sates with no guarantee of the finiteness of the internal partition function in such a case as explained above.

Finally, at the microscopic scale, one realizes that in the derivation of the entropy expression as given by Eq. (8) the degeneracy of energy levels was neglected and the expression for the entropy given by Eq. (8) in addition to the above-mentioned missing term is effectively the sum of the translational entropies of a set of various segregated compartments each of them contains what may be considered as a perfect monatomic gas at a certain specified excitation energy. The number of compartments is necessarily sufficient to contain the numbers of molecules of each kind or each excitation state at pressure P. The ignorance of the degeneracy of each level, therefore, implies that in the case of mixing (removing of the separators among compartments) there will be no entropy change in any transition between any two levels as each of these levels has <u>unit</u> a *priori* statistical weight. This fact will not be changed whether or not the excluded volume will be taken into account. Now, considering the equilibrium between any two states, the usual thermodynamic method for the determination of the distribution of atoms or molecules among the various possible states implies

$$\Delta f = -KT \ln \frac{n_j}{n_i} \tag{20}$$

where $\Delta f$ is the change of the free energy of an atom or molecule in transition between the two states. But for this simple process, in which there will be no entropy change for the above-mentioned reason, the free energy change $\Delta f$ is equal to the energy change of the atom or molecule $\Delta u$ when the particles are taken to be a perfect gas (as the interaction part is already taken into account within the entropy expression), then the ratio of the numbers in the two states is

$$\frac{n_j}{n_i} = e^{-\Delta u/KT} = e^{-(w_j - w_i)/KT} \tag{21}$$

which is the usual Boltzmann factor leading to the divergence of the internal partition function in contradiction with Fermi's results and conclusions.

## Conclusions

An important hidden inconsistency in Fermi's historic work on the probability of the quantum states is pointed out and discussed. The fact that Fermi's work has engendered many inconsistent/inaccurate models used extensively, in the literature, to calculate the equation of state of nonideal plasma systems makes it necessary to publicize this correction and clarification.

## Acknowledgments

I am indebted to Prof. T. Thiemann from the Chemistry Dept., UAE University for the time and efforts he courteously devoted in translating Fermi's original article (reference 1) from German to English.